\documentclass[11pt,a4paper,sans]{article}        
\usepackage[utf8]{inputenc}                       
\usepackage[scale=0.78]{geometry}
\usepackage{hyperref}
\usepackage{natbib, graphicx}
\usepackage{wrapfig}
\usepackage{wasysym}
\usepackage{marvosym}
\usepackage{fancyhdr}
\usepackage{multicol}
\usepackage{sectsty}
\usepackage{multirow}
\usepackage{longtable}
\pdfoutput=1

\begin{document}

{\LARGE \textsc{\bf {Planet-Host Stars Across the Galaxy in the 2040s}}} \\
\vspace{0.5cm}

{\large \textsc{\bf {Authors: M. Tsantaki (INAF - Osservatorio Astrofisico di Arcetri, Italy; maria.tsantaki@inaf.it), K. Biazzo (INAF - Osservatorio Astrofisico di Roma, Italy; katia.biazzo@inaf.it), Fatemeh Zahra Majidi (INAF-OACN, Italy), Gražina Tautvaišienė (Vilnius University, Lithuania), Innocenza Bus\`a (INAF-OACT, Italy)}}} \\





\section{Science motivation}

By the 2040s, the exoplanet field will have moved from the discovery of a few thousand planets to hundreds of thousands, thanks to Gaia DR5, TESS, PLATO, Roman, and their successors. At that stage, the key bottleneck will no longer be planet detection, but our ability to understand how planetary systems form, evolve, and diversify across different stellar and Galactic environments.

\noindent Stars and their planets form roughly at the same time from the same molecular cloud. A strong link between the composition and evolution of the star and that of its planets is therefore expected, but only partially tested. The present statistics already show that stellar iron metallicity and other elemental ratios play a crucial role in the formation of giant planets and, potentially, in shaping the architectures of planetary systems. At the same time, Galactic chemical evolution (GCE), radial migration, and birth environment imprint strong gradients in the elemental abundances and ages of stars across the Galaxy.

\noindent In the 2040s, a central science challenge will be: \textit{How do the chemical composition, age, and Galactic birth environment of stars affect the formation, evolution and fate of planetary systems?}

\noindent To address this, we need a large-scale, high-resolution spectroscopic survey of planet-host stars, spanning a broad range of Galactic environments (thin and thick disks, bulge, halo, clusters, associations), and including a well-defined control sample of non-hosts. Such a survey must deliver homogeneous stellar parameters, detailed abundance determinations, ages, and kinematics for tens of thousands of hosts, extending to the faint magnitudes probed by future missions but are beyond the reach of existing and currently planned spectroscopic facilities.

\section{Key science questions}

{\bf Planet-metallicity relation beyond giant planets.} The correlation between the frequency of giant planets and iron abundance of planet-host stars is well established, but if the same trend also holds for small planets (e.g. super-Earths, sub-Neptunes) is still under debate. Moreover, it is still not clear if other elements are as important or more important than Fe.
\\*
{\bf Role of specific elements in planet formation.} Which chemical species are critical? If $\alpha$-elements influence disk solid content and planet core mass, and volatiles shape the atmospheres and water, is there a chemical blueprint that makes certain planet architectures more likely? 
\\*
{\bf Stellar pollution from planet engulfment.} Can overabundances of refractories or volatiles be explained by planetary ingestion? If yes, what are the timescales for mixing such signatures away in the stellar convective envelope? Is there a detectable statistical link between possible abundance anomalies and planetary system dynamical instability?
\\*
{\bf Stellar abundance ratios and planet composition.} Can stellar abundance ratios (e.g. C/O, Mg/Si, Fe/Si, C/S, C/N, O/Si) be directly mapped to exoplanet composition? Do unusual stellar abundance ratios lead to “exotic” planets (e.g. carbon-rich worlds)? 
\\*
{\bf Stellar abundances/parameters versus planet properties.} Are stellar abundances of volatile or refractory elements observed for few targets correlated with the total planetary mass of the system? If so, is there also a link with system multiplicity? Is the possible correlation between planetary orbital eccentricity, stellar iron abundance, and planet density observed for giant planets genuine, or could it be the result of unrecognized biases? If genuine, does it also hold for rocky-planet-host stars? Are planet–abundance correlations dependent on stellar age?
\\*
{\bf Abundance patterns as fingerprints of planet formation.} It is still debated whether planet-host stars have systematic abundance anomalies compared to non-hosts. Furthermore, it remains unclear if the refractory depletion seen in the Sun, when compared to solar twins without planets, is due to terrestrial planet formation or to biases in the GCE at the time and place the Sun formed.
\\*
{\bf Chemical evolution versus planet formation effects.} We will have the possibility to separate the chemical trends due to the Galactic evolution from those due to the actual planet formation processes. Does GCE affect the formation of certain planets in different epochs in our Galaxy?
\\*
{\bf Star-planet interaction triggering.} In high resolution, we will be able to infer the geometric and energetic configuration of the stellar atmosphere (from the photosphere to the corona), which is decisive in triggering star-planet interactions such as auroral-radio-emission.
\\*
{\bf Planet photo-evaporation through stellar chromospheric activity.} The use of several activity indicators (e.g. Ca\,H\&K lines, Na\,D lines, H$\alpha$, and Ca infrared triplet) for tens of thousands of planet host stars will enable investigations of the effects of star-planet interactions and magnetic enhancement, which can have significant implications for photo-evaporation of young planetary atmospheres.
\\*
{\bf Non-LTE modelling to disentangle short-term magnetic activity variability from the effects of exoplanets in transmission spectroscopy.} Transmission spectroscopy of transiting exoplanets has proven to be one of the most effective techniques for atmospheric characterization. However, the accuracy of the inferred planetary spectra is strongly affected by the host star’s magnetic activity, which can introduce variability that either mimics or masks genuine planetary atmospheric signals. This stellar noise represents one of the major limitations to transmission spectroscopy, particularly for magnetically active late-type stars. We will explore the correlations among different diagnostics relevant to the study of planetary atmospheres, by performing NLTE radiative transfer calculations on a grid of extended M-dwarf atmospheric models characterized by different levels of magnetic activity. The resulting relationships can be used to disentangle the effects of short-term magnetic variability from those caused by the planet’s atmosphere. 

\section{Why a new facility is needed beyond the 2030s}

\begin{figure*}
  \centering
  \includegraphics[width=0.5\linewidth]{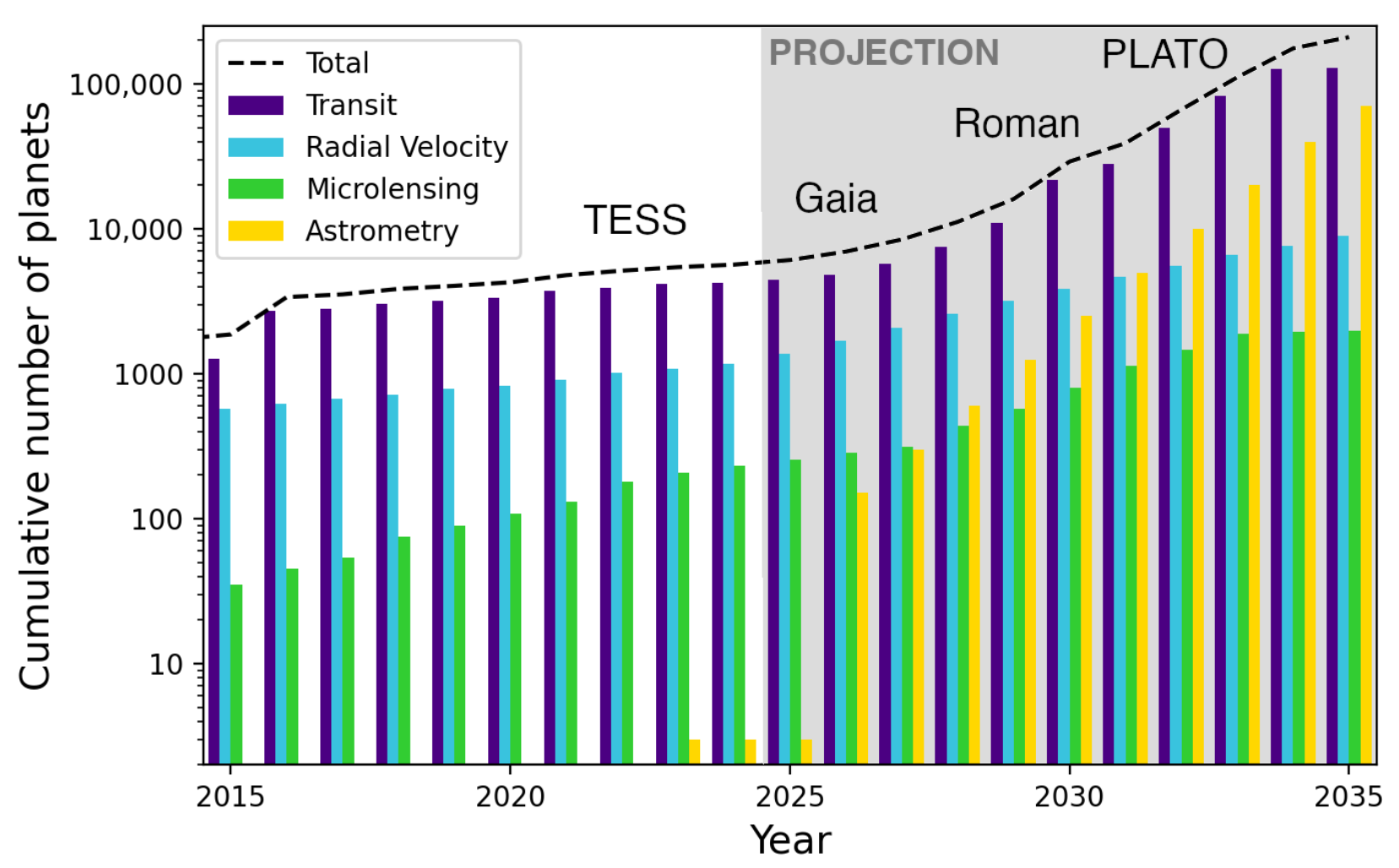}  \\ 
  \caption{Number of planets detected up to 2025 and a projection of the expected planets to be discovered in the future from J. Christiansen et al. (2025).}
  \label{exoplanets}
\end{figure*}

By the late 2030s, several powerful facilities will be in place both from ground and space. Space missions (Gaia, TESS, PLATO, Roman) will provide planet detections for $\sim$200\,000 systems (Fig.~\ref{exoplanets}). Medium-resolution, multiplexed spectrographs (4MOST, WEAVE, MOONS) and ELT instruments (MOSAIC, ANDES) will deliver valuable spectroscopy, but with strong limitations in either spectral resolution, telescope size, field of view, or multiplexing. However, no facility currently existing or planned can deliver at the same time:
\begin{itemize}
    \item High resolution (R$>$40,000) and high multiplexing on a large (10--12\,m) aperture over a wide field of view, enabling precise abundances for tens of thousands of faint planet hosts.
    \item Simultaneous coverage of key diagnostics (e.g. C, N, O, refractories, n-capture elements, activity indicators) at sufficient resolution to measure weak and blended lines in cool dwarfs. 
    \item Depth and scale to reach: faint Roman transit hosts in the inner Galaxy ($13<G<20$ mag), M dwarfs from PLATO and TESS at $G>13$.
    \item A homogeneous and statistical control sample of non-planet-host stars in the same Galactic environments, observed with identical setups, to disentangle GCE from planet-formation signatures.
\end{itemize}

Current and near-term ESO MOS facilities (e.g., FLAMES, 4MOST, MOONS, ELT/MOSAIC) typically provide R$\sim$20,000 in multiplex mode or are limited in field of view and multiplex at higher resolution. At R$\sim$20,000, many key abundance diagnostics, particularly for C, N, O, and for complex cool-star spectra, become too blended or too shallow to reach the less than 0.05\,dex precision required to address the scientific questions above, especially for faint targets. Thus, the science case outlined here requires a new, wide-field, high-multiplex, high-resolution spectroscopic facility on a large-aperture telescope that is not expected to exist before the 2040s. Concepts such as a 10–12\,m class Wide-field Spectroscopic Telescope (WST) illustrate the type of facility that would be transformative for this science.

\section{Technology requirements}

The science case implies the following capability targets:
\begin{itemize}
    \item High-resolution MOS mode: R$\sim$40,000 over broad optical coverage, with several thousand fibres per pointing, on a 10~m aperture to reach $G\sim$21--22 in reasonable exposure times.
    \item Complementary low/medium-resolution MOS: For activity indicators, time-domain monitoring, and target vetting, including planet-candidate hosts.
    \item NIR coverage (optional): Access to NIR lines of key elements (e.g. additional CNO diagnostics, P, K, S, molecular bands in cool dwarfs) to improve abundances and ensure consistency with exoplanet atmosphere studies, many of which will be NIR-based. In the context of stellar magnetic activity diagnostics, the Ca II infrared triplet (8498, 8542, 8662~\AA) offers significant advantages over the traditional Ca II H \& K (3969--3934~\AA) lines. The IRT region lies in a portion of the spectrum with a well-defined continuum, which facilitates a more accurate subtraction of the photospheric contribution---an essential step for quantifying chromospheric emission. 
    \item IFU(s): For crowded regions (bulge, cluster cores) to resolve individual stars and obtain clean spectra in environments where planet formation and survival may be strongly environment-dependent.
\end{itemize}

\section{References}
{\small
Adibekyan et al. (2019, Geosciences, 9, 105), Adibekyan et al. (2021, Science, 374, 330), Biazzo et al. (2022, A\&A, 664, A161), Boettner et al. (2024, A\&A, 692, A150), Christiansen, et al. (2025, The Planetary Science Journal, 6), Dorn et al. (2015, A\&A, 577, A83), Magrini et al. (2022, A\&A, 663, A161), Spina et al. (2021, NatAst, 5, 1163), Sun et al. (2025, ApJ, 978, 107), Swastik et al. (2022, AJ, 164, 60), Tautvaisiene et al. (2022, ApJS 259, 80), Turrini et al. (2021, ApJ, 909, 40), Valencia et al. (2013, ApJL, 775, L4), Weeks et al. (2025, MNRAS, 539, 405)}

\end{document}